\documentclass{PoS}
\usepackage{amsmath}
\usepackage{graphicx}
\title{Nuclear forces in the parity odd sector and the LS forces}

\ShortTitle{Nuclear forces in the odd parity sector and the LS forces}
\author{\speaker{Keiko Murano}\\
        RIKEN Nishina Center, RIKEN, Wako 351-0198, Japan\\
        E-mail: \email{murano@riken.jp}}

	\author{for HAL QCD Collaboration
	}

\abstract{
In  this  paper, we  report our first  attempt  at determining  NN
potentials in the parity odd sector including the spin-orbit force in lattice QCD, 
employing the method to extract successfully parity even NN potentials
from  Nambu-Bethe-Salpeter (NBS)  wave functions  through  the  Schr\"odinger  equation.
Using  $N_f=2$ CP-PACS  gauge configurations  on a $16^3 \times 32$ lattice at $a\simeq 0.16$ fm and $m_\pi\simeq 1.1$  GeV, we calculate
central,  tensor and spin-orbit potentials  in the parity odd sector.
Although statistical  errors are still large,  we observe
that the qualitative features  of these  potentials roughly agree
with those of phenomenological potentials.
}
\FullConference{ The XXIX International Symposium on Lattice Field Theory - Lattice 2011\\
July 10-16, 2011\\
Squaw Valley, Lake Tahoe, California}

\begin{document}

\section{Introduction}
The nucleon-nucleon (NN) potential  is one of the fundamental quantity
to  study  various properties  of  atomic  nuclei.   In the  past  few
decades,                based                on                various
models\cite{Machleidt:2000ge,Wiringa:1994wb,Stoks:1994wp}    or    the
chiral effective field theory\cite{Epelbaum:2008ga}, several realistic
NN  potentials have been  constructed to  reproduce the  NN scattering
phases of more than 4000 data points with $\chi^2/N_{\rm dof} \sim 1$.
On  the  other  hand, quite  recently,  a  new  method to  extract  NN
potential  in  QCD  has  been  proposed\cite{Ishii:2006ec}.
For  the moment,  the  method  has been  successfully  applied to  the
potentials  in the  parity even  sector at  the leading  order  of the
derivative   expansion,  i.e.,  the   central  force   (  $V^{(+)}_0$,
$V^{(+)}_\sigma$) and the tensor  force ($V^{(+)}_T$) are obtained for
various      cases       including      NN,      YN,       and      YY
systems\cite{Nemura:2008sp,Inoue:2010es,Inoue:2010hs}     (where     Y
represents a hyperon) and meson-baryon system\cite{arXiv:1111.2663}.
In  addition,  the  method  is  recently extended  to  investigate  3N
forces\cite{Doi:2011gq}.
In  contrast, the  method has  not yet  applied to  the forces  in the
parity odd  sector as well as  the spin-orbital force.
They are  needed for  the complete determination  of NN  potentials at
next-to-leading order (NLO).
In  particular,  the  spin-orbit  force,  even though  it  appears  at
NLO  of a derivative expansion, is  important, as is
indicated  by  an  analysis  of  phase shift\cite{Bohr,352344}. 
It is  also important  in explaining the  ls-splitting of  the (hyper)
nuclear spectra and the magic  numbers in nuclei, i.e., it induces the
one-body  spin-orbit  term  in the  average  single-particle  nuclear
potential\cite{Bohr,44721,Jensen}.

The purpose of  this paper is to extract potentials  in the parity odd
sector including  the spin-orbit  force for the  NN system.   For this
purpose, we construct NBS  wave functions which include higher angular
momenta, in order to determine the central force, the tensor force and
the  spin-orbit   force  in   the  parity  odd   sector  ($V_0^{(-)}$,
$V_\sigma^{(-)}$,  $V_T^{(-)}$, $V_{LS}^{(-)}$).  Our  calculation has
been  made by  using  $Nf=2$  CP-PACS configurations  on  a $T  \times
L^3=16^3  \times 32$  lattice at  $\beta=1.95$ ($a\simeq 0.16$  fm) and
$\kappa=0.1375$ ($m_\pi\simeq 1100$ MeV).

\section{Extraction of potentials}
\subsection{Definition of potentials}
In the method,
the NN potential is constructed form Nambu-Bethe-Salpeter (NBS) wave
function in the center of mass (CM) frame,  defined by
\begin{eqnarray}
\phi_{\alpha,\beta}({\bf r};E)\equiv \langle 0 | p_\alpha({\bf x})
 n_\beta({\bf y}) | p(+{\bf k})n(-{\bf k})\rangle, \quad ({\bf r}\equiv
 {\bf x}-{\bf y}) \label{eq:NBS}
\end{eqnarray}
where $p_\alpha$ and  $n_\beta$ denote local composite nucleon operators with
spinor indices $\alpha$, $\beta$, and
$E$ is kinetic energy related to the relative momentum $k = \vert{\bf k}\vert$ as
$E=2\sqrt{k^2+m_N^2}-2m_N(\sim k^2/m_N)$ with $m_N$ being the nucleon mass.
We define NN potentials from the NBS wave function below the inelastic threshold
through the following Schr\"odinger type equation \cite{Aoki:2009ji,Aoki:2008hh}
\begin{eqnarray}
	\left(E+\frac{\nabla^2}{m_N}  \right) \phi(\vec{r};E) = \left[ P^+ V^{(+)} (r) + P^- V^{(-)}(r) \right] \phi(\vec{r};E),
	\label{eq:pot1}
\end{eqnarray}
	where $P^{+}$ ($P^{-}$) denotes a projection operator for parity even (odd) with the corresponding potential $V^{(+)}$, ($V^{(-)}$), which is decomposed as
\begin{eqnarray}
	V^{(\pm)}(r) &=& \left[ V^{(\pm)}_0(r) + V^{(\pm)}_{\sigma}(r) \vec{\sigma}_1\cdot\vec{\sigma}_2 + V^{(\pm)}_T(r) S_{12} + V^{(\pm)}_{LS}(r) \vec{L}\cdot\vec{S} +({\rm NNLO}) \right]  \label{eq:pot2}  
\end{eqnarray}
with the central and spin-dependent central forces $V_0^{(\pm)}$ and $V_\sigma^{(\pm)}$, 
the tensor force $V_T^{\pm}$ , and the spin-orbit force $V_{LS}^{(\pm)}$.  While $V_0^{(\pm)}$,
$V_\sigma^{(\pm)}$ and $V_T^{(\pm)}$ are of leading order(LO) in the derivative
expansion of  non-local potentials, $V_{LS}^{\pm}$  appears at next-to-leading order(NLO). 
Once the above NBS wave functions are calculated in lattice QCD simulations, 
potentials can be extracted by solving eq.(\ref{eq:pot1}).
\subsection{Spin-singlet potentials}
For the spin-singlet and parity odd sector,
the Schr\"odinger type equation eq.(\ref{eq:pot1}) is reduces to
\begin{eqnarray}
	\left(E+\frac{\nabla^2}{m_N}\right)
	 \phi^{(-)}_{S=0}(\vec{r};E) =  Vc^{(-)}_{S=0}(r)
	 \phi^{(-)}_{S=0}(\vec{r};E),  
\end{eqnarray}
where $\phi^{(-)}_{S=0}(\vec{r};E) = P^{(-)} P^{S=0} \phi(\vec{r};E)$ with the spin-projection operator $P^S$  and $Vc^{(-)}_{S=0}(r)=V_0^-(r)-3V_\sigma^-(r)$ denotes the central potential in the spin-singlet channel. From this equation,  we then obtain 
\begin{eqnarray}
 Vc_{S=0}^{(-)}(r)=E+\frac{1}{m_N}\frac{\nabla^2 \phi({\bf
  r};E)}{\phi({\bf r};E)}. \label{eq:Vc_S0}
\end{eqnarray}

\subsection{Spin-triplet potentials including spin-orbit force}
For the spin-triplet and parity odd sector, the Schr\"odinger type equation
eq.(\ref{eq:pot1}) reads
\begin{eqnarray}
&&	\left(E+\frac{\nabla^2}{m_N} \right) \phi^{(-)}_{S=1}(\vec{r}) =
	 \left[ Vc^{(-)}_{S=1}(r) + V_T^{(-)}(r) \ S_{12} + V_{LS}^{(-)}
	  (r) \ \vec{L}\cdot\vec{S} \right] \phi^{(-)}_{S=1}(\vec{r}),
	 \label{eq:SE_S1}.
\end{eqnarray}
Where $\phi^{(-)}_{S=1}({\bf r}) =  P^{(-)}P^{S=1}\phi({\bf r};E)$
and $Vc^{(-)}_{S=1}(r) = V_0^-(r) + V_\sigma^-(r)$. Note that the isospin of these $V_0^-$ and $V_\sigma^-$ is different from that of the spin singlet sector. 

In order to determine the spin-orbit potential for parity odd sector,
we consider three linear independent set of NBS wave functions,
$\phi^{X_i}$ with $i=1,2,3$,  constructed by
\begin{eqnarray}
	\phi^{X_i}(\vec{r})  = P^{X_i} P^{S=1} P^{(-)} \phi(\vec{r};E).
\end{eqnarray}
where $P^{X_i}$ denotes a projection operator to state $X_i$, labeled by the orbital
angular momentum $L$ and the total spin $J$.
Eq.~(\ref{eq:SE_S1} is then decomposed into three independent equations as 
\begin{eqnarray}
	\left(E+\frac{\nabla^2}{m_N} \right) \phi^{X_i}(\vec{r}) = P^{X_i} \left[ V_C^{(-)}(r) + V_T^{(-)}(r) \ S_{12} + V_{LS}^{(-)} (r) \ \vec{L}\cdot\vec{S} \right] \phi(\vec{r}) .
	 \label{eq:simul}
\end{eqnarray}
By solving these equations,
potentials $V_C^{(-)}$, $V_T^{(-)}$ and $V_{LS}^{(-)}$ can be obtained as
\begin{eqnarray}
	 \left(
		\begin{array}{c}
			V_C^{(-)}(r)-E \\
			V_T^{(-)}(r) \\
	  	V_{LS}^{(-)}(r)
		\end{array}
	\right)
	=
		 \ M(\vec{r})^{-1}
		\left(
		\begin{array}{c}
			\nabla^2/m_N \ \phi^{X_1}(\vec{r}) \\
		          \nabla^2/m_N \ \phi^{X_2}(\vec{r}) \\
			\nabla^2/m_N \ \phi^{X_3}(\vec{r}) 
		\end{array}
		\right), \label{eq:solve}
\end{eqnarray}
where $M(\vec{r})$ denotes $3\times3$ matrix defined by
\begin{eqnarray}
	M(\vec{r}) =
\left(
	\begin{array}{ccc}
		\phi^{X_1}(\vec{r})  \ \ & P^{X_1} S_{12} \phi(\vec{r}) \ \  & P^{X_1} \vec{L}\cdot\vec{S} \phi(\vec{r}) \\
				\phi^{X_2}(\vec{r})  \ \ & P^{X_2} S_{12} \phi(\vec{r}) \ \  & P^{X_2} \vec{L}\cdot\vec{S} \phi(\vec{r}) \\
						\phi^{X_3}(\vec{r})  \ \ & P^{X_3} S_{12} \phi(\vec{r}) \ \  & P^{X_3} \vec{L}\cdot\vec{S} \phi(\vec{r})
	\end{array}
	\right).
\end{eqnarray}

\section{Construction of NBS wave functions in lattice QCD}
The  NBS wave  function  is obtained  from
a 4-point nucleon correlation function on the lattice as
\begin{eqnarray}
	G_{\alpha,\beta}(\vec{x}-\vec{y},t-t_0;\mathcal{J}) \equiv \frac{1}{L^3} \sum_{\bf r} \langle 0 | T\left[p_\alpha(\vec{x}+\vec{r},t) n_\beta(\vec{y}+\vec{r},t)\mathcal{J}^{J=\tilde{J},S=\tilde{S}}(t_0)\right] | 0 \rangle
\label{eq:4-point}	
	\\
 \simeq
  \phi^{J=\tilde{J},S=\tilde{S}}_{\alpha,\beta}(\vec{x}-\vec{y};E_0)
  \langle p(+{\bf k}_0)n(-{\bf k}_0) | \mathcal{J}^{J=\tilde{J},S=\tilde{S}}(0)|0\rangle e^{-E_0 (t-t_0)}, \quad t-t_0 \gg 1,
\end{eqnarray}
where  the summation over  ${\bf r}$  is performed  to select  the two
nucleon system with vanishing total spatial
momentum.
For  $p(x)$  and  $n(y)$,  we  employ the  following  local  composite
operators
\begin{eqnarray}
 p(x)\equiv \epsilon_{abc} (u_a^t(x) \gamma_5 d_b(x)) u_c(x), \quad
  n(x) \equiv \epsilon_{abc} (u^t_a(x) C\gamma_5 d_b(x)) d_c(x),
\end{eqnarray}
where $a$, $b$ and $c$ denote color indices,
while 
$\mathcal{J}^{J=\tilde{J},S=\tilde{S}}$  denotes a  two-nucleon source
operator with a definite  total spin $S=\tilde{S}(\tilde{S}=0, 1)$ and
a definite total  "angular momentum" $J=\tilde{J}$,
which is indeed  an  irreducible  representation  of the  cubic group as $\tilde{J}=A_1, A_2,
E,  T_1,  T_2$.

For such two-nucleon  sources, we take two-nucleon momentum
wall sources defined by
\begin{equation}
  \mathcal{J}_{\alpha\beta}(f) \equiv \bar P_{\alpha}(f) \bar N_{\beta}(f^*),
\end{equation}
where
\begin{eqnarray}
  \bar P_{\alpha}(f)
  &\equiv&
  \sum_{{\bf x}_1,{\bf x}_2,{\bf x}_3}
  \epsilon_{abc}
  \left(
    \bar{u}_{a}({\bf x}_1) C\gamma_5 \bar{d}_{b}({\bf x}_2)
  \right)
  \bar{u}_{c,\alpha}({\bf x}_3) f({\bf x}_3)
  \\\nonumber
  \bar N_{\beta}(f)
  &\equiv&
  \sum_{{\bf x}_1,{\bf x}_2,{\bf x}_3}
  \epsilon_{abc}
  \left(
    \bar{u}_{a}({\bf x}_1) C\gamma_5 \bar{d}_{b}({\bf x}_2)
  \right)
  \bar{d}_{c,\beta}({\bf x}_3) f({\bf x}_3)
\end{eqnarray}
with $f$  being one of the  following source functions,  each of which
corresponds to a plain wave parallel to one of  spatial coordinate axes as
\begin{eqnarray}
 f^{(0)}({\bf r}) \equiv \exp(-2\pi i x/L),
 \quad
 f^{(1)}({\bf r}) \equiv \exp(-2\pi i y/L),
 \quad
 f^{(2)}({\bf r}) \equiv \exp(-2\pi i z/L),
 \nonumber \\
 f^{(3)}({\bf r}) \equiv \exp(+2\pi i x/L),
 \quad
 f^{(4)}({\bf r}) \equiv \exp(+2\pi i y/L),
 \quad
 f^{(5)}({\bf r}) \equiv \exp(+2\pi i z/L). 
\end{eqnarray}
Note that  an element $g$ in the  cubic group $O$  with  24 elements acts  on these plane
waves as
\begin{eqnarray}
 f^{(i)} \mapsto  \sum_{j}  U_{ij}(g)  f^{(j)},
\end{eqnarray}  
where the $6\times 6$ representation matrix $U(g)$
are generated by  representation matrices for ${\bf c}_y$ and
${\bf c}_z$,  which are  the rotations by 90 degree around y and  z axes, respectively, and are explicitly given by
\begin{equation}
  U({\bf c}_y)
  \equiv
  \left[
    \begin{array}{cccccc}
      0 & 0 & 1 &   0 & 0 & 0 \\
      0 & 1 & 0 &   0 & 0 & 0 \\
      0 & 0 & 0 &   1 & 0 & 0 \\
      0 & 0 & 0 &   0 & 0 & 1 \\
      0 & 0 & 0 &   0 & 1 & 0 \\
      1 & 0 & 0 &   0 & 0 & 0
    \end{array}
  \right],
  \quad
  U({\bf c}_z)
  \equiv
  \left[
    \begin{array}{cccccc}
      0 & 0 & 0 &   0 & 1 & 0 \\
      1 & 0 & 0 &   0 & 0 & 0 \\
      0 & 0 & 1 &   0 & 0 & 0 \\
      0 & 1 & 0 &   0 & 0 & 0 \\
      0 & 0 & 0 &   1 & 0 & 0 \\
      0 & 0 & 0 &   0 & 0 & 1
    \end{array}
  \right].
\end{equation}
Note also that the spatial reflection  for these source functions is
represented by the complex conjugation.
By a cubic group analysis, the orbital part of this momentum wall
source is decomposed into $A_1^+ \oplus E^+ \oplus T_1^-$.
Therefore, for  the parity odd sector, $J^P$
we  can   access  is   $(L=T_1^-)\otimes   (S=A_1)=T_1^-$  for
the spin-singlet sector and $(L=T_1^-)  \otimes (S=T_1) = A_1^- \oplus E^-
\oplus T_1^- \oplus T_2^-$ for the spin-triplet sector.

The wall source with the definite total angular momentum  is now constructed as
\begin{equation}
  \mathcal{J}^{J=\tilde{J}}_{\alpha\beta}(f^{(i)})
  \equiv
  \frac{d^{(\tilde{J})}}{24}
  \sum_{g\in O}
  \chi^{(\tilde J)}(g^{-1})
  U_{ij}(g)
  \mathcal{J}_{\alpha'\beta'}(f^{(j)})
  S^{-1}_{\alpha'\alpha}(g^{-1})
  S^{-1}_{\beta'\beta}  (g^{-1})
\end{equation}
where   $d^{(\tilde{J})}$  and   $\chi^{(\tilde  J)}(g)$   denote  the
dimension  and  the   character  for  the  irreducible  representation
$\tilde{J}$,  respectively.  
Hereafter the Dirac indices $\alpha, \beta$ are restricted to upper components (in the Dirac representation).
The total spin $S$ is projected  by the spin projection operator $P^{(S)}$
as
\begin{equation}
  \mathcal{J}^{J=\tilde{J},S={\tilde{S}}}_{\alpha\beta}(f^{(i)})
  \equiv
   P^{(\tilde{S})}_{\alpha\beta,\gamma\delta}
  \mathcal{J}^{J=\tilde{J}}_{\gamma\delta}(f^{(i)}),
\end{equation}
$P^{(S=0)}\equiv (1-{\bf \sigma}_1\cdot{\bf \sigma}_2)/4$ and
$P^{(S=1)}\equiv (3 + {\bf \sigma}_1\cdot{\bf \sigma}_2)/4$. 
Finally,  the parity projection is defined by
\begin{equation}
  P^{(\pm)}\cdot\mathcal{J}_{\alpha\beta}(f^{(i)})
  =
  \frac1{2}
  \left(
  \mathcal{J}_{\alpha\beta}(f^{(i)})
  \pm
  \mathcal{J}_{\alpha\beta}(f^{(i)*})    
  \right).
\end{equation}

Although the  orbital part for our two-nucleon source  operators in the
parity odd system consists  of $L=T_1$ only, other orbital  components of the NBS wave function at the sink are induced through the effect of the tensor force in the case of the spin-triplet sector. 
 In this paper, for simplicity,  we consider only an 
$L=T_1$ component of  the NBS wave function as
\begin{eqnarray}
  \phi^{J=\tilde{J}, S=\tilde{S}}_{L=T_1}({\bf r};E_0)
  \equiv
  \sum_{g\in O}
  \chi^{(T_1)}(g)^*
  \phi^{J=\tilde{J},S=\tilde{S}}(R(g)
  {\bf r};E_0).
\end{eqnarray}
where
$R(g)$  denotes the  rotation matrix  for a  cubic group  element $g$,
which rotates only the orbital components of the NBS wave functions as
${\bf x} \mapsto {\bf x}' \equiv R(g){\bf x}$.
\section{Lattice setup}
Our  calculation  is performed  on  a  set  of $N_f=2$  dynamical  QCD
configurations     generated    by    the     CP-PACS    Collaboration
on a $16^3\times 32$ lattice\cite{hep-lat/0105015}, who employs
the $O(a)$-improved   Wilson   quark    action   with   $C_{SW}=1.53$   at
$\kappa_{ud}=0.1375$ and the RG improved gauge action (Iwasaki action)
at $\beta=1.95$.   This set of  parameters corresponds to  the lattice
spacing $a=0.1555$ fm and pion mass $m_\pi=1136$ MeV( nucleon mass
$m_N=2165$ MeV).
The 4-point nucleon correlation function eq.~(\ref{eq:4-point}) is calculated
with the periodic and  the Dirichlet boundary conditions along the
spatial and  the temporal directions, respectively.
The spatial  momentum of  each nucleon is  discretized as  $k_i \simeq
2\pi  n_i/L(n_i\in \mathbb{Z})$  in the  periodic BC.
Since $k  = 0$ state is  forbidden in the parity odd  system, the ground
state  energy with  parity  odd  system is  not  $E=0$ but  $E\simeq
(2\pi/L)^2/m_N$.
In  our analysis, we thus employ a free value $E=(2\pi/L)^2/m_N=115$ MeV for an energy $E$ in eq.~(\ref{eq:solve}).

\section{Numerical results}
\begin{figure}
\begin{center}
\scalebox{0.5}{\includegraphics[]{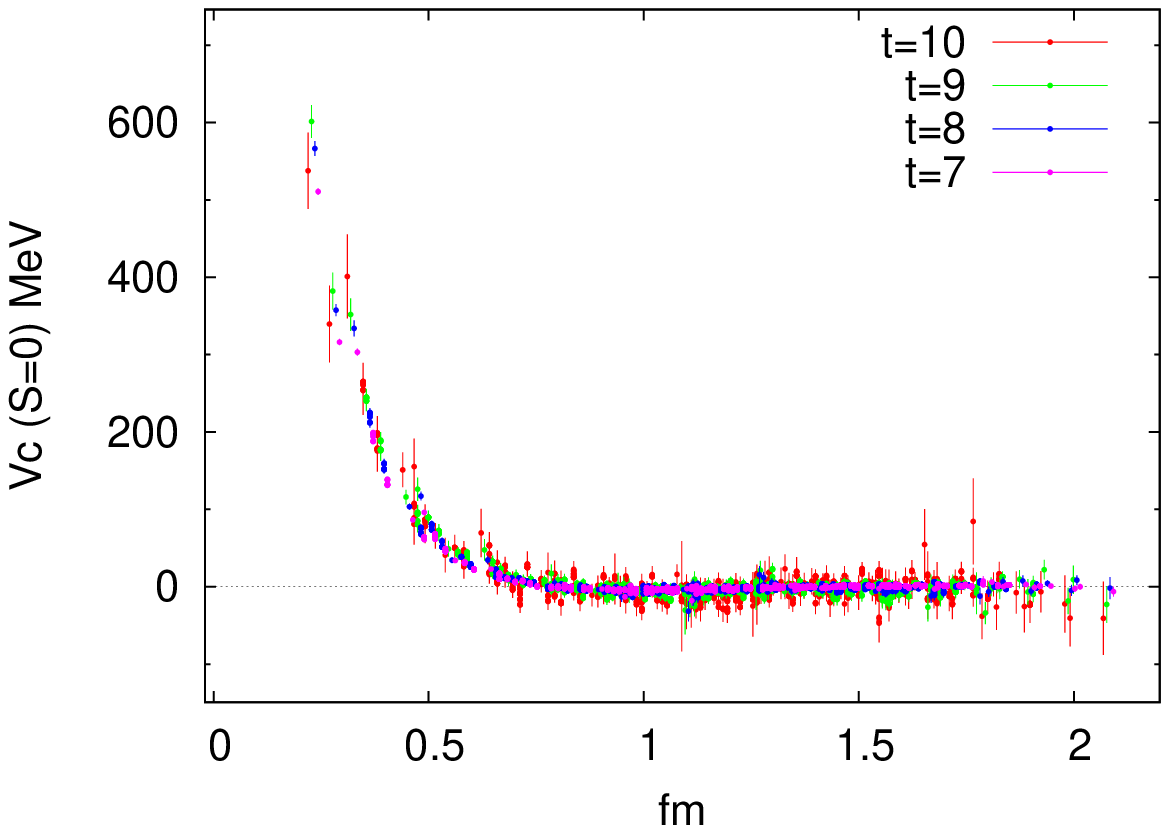}}
\scalebox{0.5}{\includegraphics[]{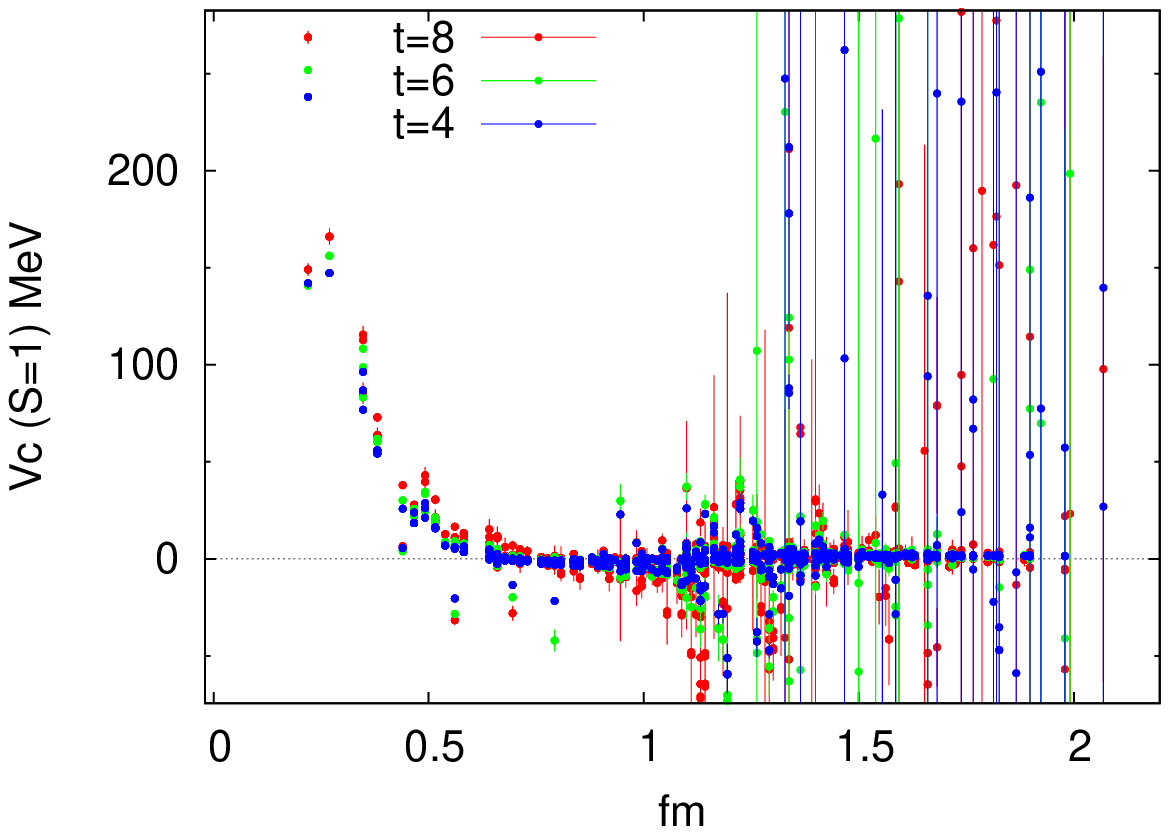}}
\end{center}
\caption{(Left) The central potential in the spin-singlet and parity odd
 sector, calculated from $^1P_1$ NBS wave function. Data at $t-t_0=7,8,9,10$ are
 simultaneously plotted.
 (Right)The central potentials in the spin-triplet and parity odd
 sector, calculated from $^3P_0$, $^3P_1$ and $^3P_2$ NBS wave
 functions. Data at $t-t_0=4,6,8$ are
 simultaneously plotted.}
\label{fig:Vc_both}
\end{figure}

For the spin singlet sector,
we calculate the $J^P  =T_1^-$  NBS wave function with $L=T_1$,
$\phi^{J=T_1,  S=0}_{L=T_1}$,  whose dominant component corresponds  to  the $^1P_1$ NBS  wave
function.   Using   eq.(\ref{eq:Vc_S0}),  we  obtain  the  central
potential in this  channel.  Preliminary results at $t-t_0=7-10$ are  plotted  in  Fig.~\ref{fig:Vc_both} (Left).
We observe that  the central potential has a strong repulsion at  short distance ($r \le
0.5$ fm) without attraction at longer distance. 

\begin{figure}
\scalebox{0.5}{\includegraphics[]{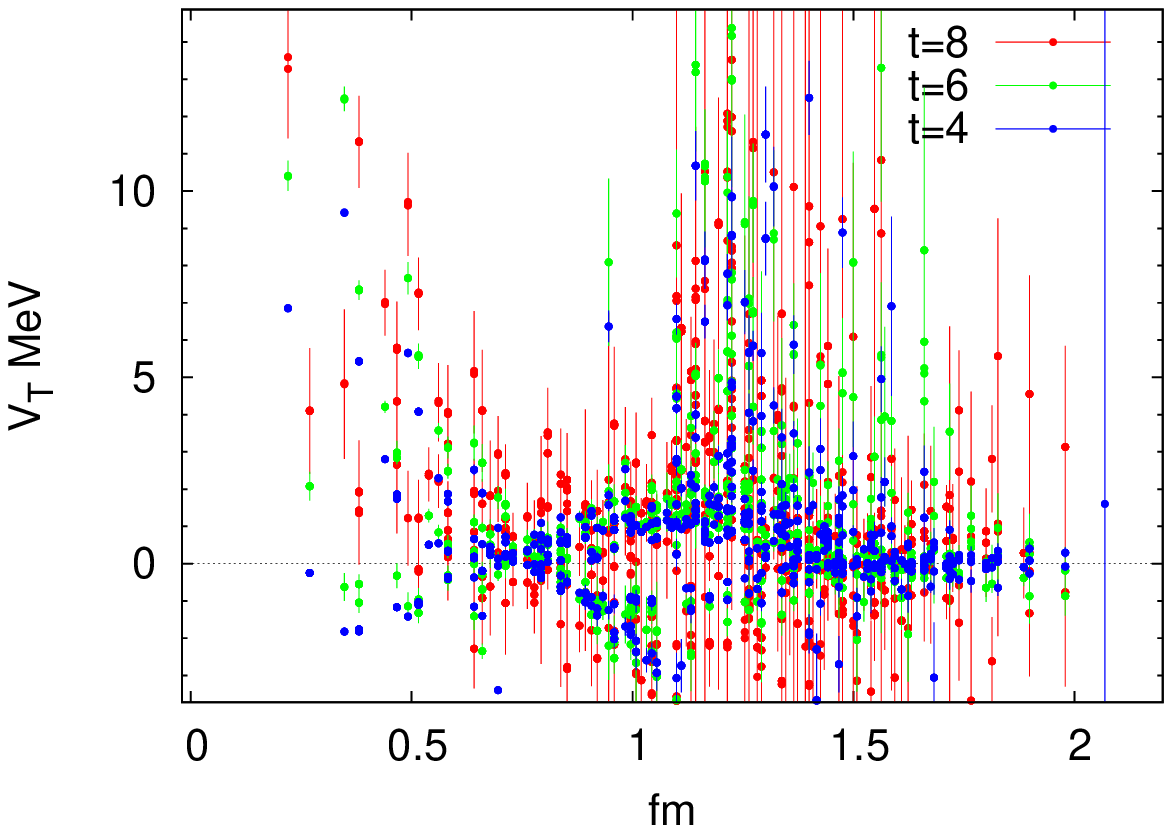}}
\scalebox{0.5}{\includegraphics[]{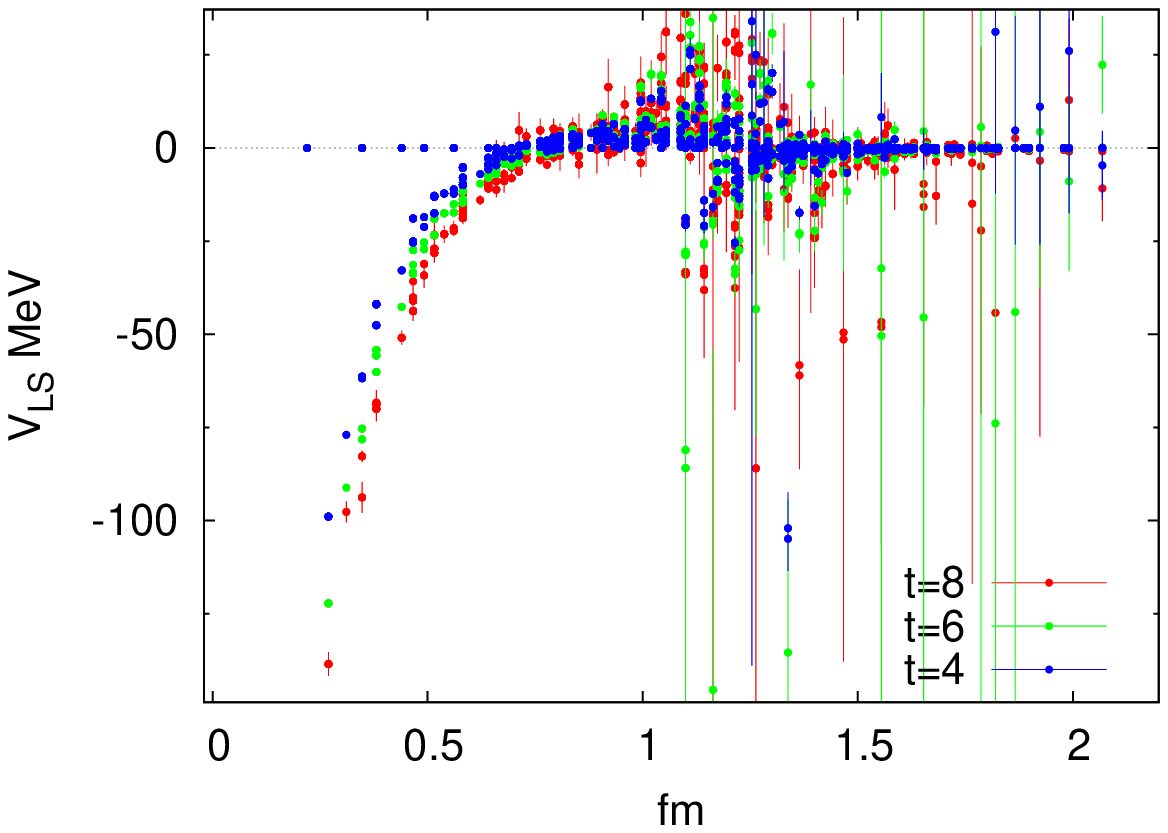}}
\caption{Potentials with spin-triplet state in parity odd
 sector, calculated from $^3P_0$, $^3P_1$ and $^3P_2$ NBS wave
 functions. Data at $t-t_0=4,6,8$ are
 simultaneously plotted.
 (Left) The tensor potential.
 (Right) The spin-orbit potential.}
\label{fig:VLS}
\end{figure}

For the spin triplet sectors,
we   calculate    three   NBS   wave    functions   $\phi^{J=\tilde{J},
  S=1}_{L=\tilde{T_1}}$  with  $\tilde{J}=A_1$,$T_1$  and  $E$,  whose dominant components
correspond to  $^3P_0$, $^3P_1$  and $^3P_2$, respectively.   Using
these  NBS wave  functions,  we extract  spin-triplet potentials  in
parity odd  sector.
The central potential $Vc^{(-)}_{S=1}(r)$ is shown in Fig.~\ref{fig:Vc_both} (Right).  Moreover
the tensor  potential  $V_T^{-}(r)$ and the  spin-orbit force  $V_{LS}^{(-)}(r)$ are given in 
the left and right panels of Fig.~\ref{fig:VLS}, respectively.
Large  fluctuations  at
$r\simeq  1.2$ fm  are  caused by  spatial boundaries  at $L/2=8$,  which breaks  the
rotational symmetry, while
large statistical  errors of the tensor force  in the parity odd sector
may be  mainly caused by  the fact that  its magnitude is  significantly smaller
(few MeV) than other  potentials (few-hundred MeV).
Although systematic and statistical uncertainties are still large in
our  preliminary results, we  observe the following qualitative features.
(1) $V_C$ is repulsive at all distance with a repulsive core at short distance.
 (2)  $V_T$ is positive and quite small.  (3)
  $V_{LS}$ is large and negative.
These  features  qualitatively agree  with those of phenomenological potentials\cite{Wiringa:1994wb}.

\section{Conclusions}
In  this  paper, we  have  made a first  attempt  at determining  NN
potentials in the parity odd sector including the spin-orbit force.
Using  $N_f=2$ CP-PACS  gauge configurations  on a $16^3 \times 32$ lattice at $a\simeq 0.16$ fm and $m_\pi\simeq 1100$  MeV, we  have reported our preliminary results on
central forces,  the tensor force  and the spin-orbit force  in the parity odd sector.
Although statistical  errors are still rather large,  we have observed
that the qualitative behaviors  of these  potentials roughly agree
with those of phenomenological potentials.

Numerical calculations are performed on University of Tsukuba Supercomputer
system (T2K).
This work is supported by the Grant-in-Aid for Scientific Research on Innovative Areas(No.2004: 20105001, 20105003) and for Scientific Research(C) 23540321.
We are grateful for authors and maintainer of CPS++\cite{cps} a modified version of which is used for this work.

\end{document}